# Impact of Interlayer and Ferroelectric Materials on Charge Trapping during Endurance Fatigue of FeFET with TiN/Hf$_x$Zr$_{1-x}$O$_2$/interlayer/Si (MFIS) Gate Structure

Fengbin Tian, Shujing Zhao, Hao Xu, Jinjuan Xiang, Tingting Li, Wenjuan Xiong, Jiahui Duan, Junshuai Chai, Kai Han, Xiaolei Wang, Wenwu Wang, and Tianchun Ye

*Abstract*—We study the impact of different interlayers and ferroelectric materials on charge trapping during the endurance fatigue of Si FeFET with TiN/Hf$_x$Zr$_{1-x}$O$_2$/interlayer/Si (MFIS) gate stack. We have fabricated FeFET devices with different interlayers (SiO$_2$ or SiON) and Hf$_x$Zr$_{1-x}$O$_2$ materials (x=0.75, 0.6, 0.5), and directly extracted the charge trapping during endurance fatigue. We find that: 1) The introduction of the N element in the interlayer suppresses charge trapping and defect generation, and improves the endurance characteristics. 2) As the spontaneous polarization ($P_s$) of the Hf$_x$Zr$_{1-x}$O$_2$ decreases from 25.9 µC/cm$^2$ (Hf$_{0.5}$Zr$_{0.5}$O$_2$) to 20.3 µC/cm$^2$ (Hf$_{0.6}$Zr$_{0.4}$O$_2$), the charge trapping behavior decreases, resulting in the slow degradation rate of memory window (MW) during program/erase cycling; in addition, when the $P_s$ further decreases to 8.1 µC/cm$^2$ (Hf$_{0.75}$Zr$_{0.25}$O$_2$), the initial MW nearly disappears (only ~0.02 V). Thus, the reduction of $P_s$ could improve endurance characteristics. On the contract, it can also reduce the MW. Our work helps design the MFIS gate stack to improve endurance characteristics.

## I. Introduction

Hafnium-based FeFET is perused due to its excellent characteristics, such as > 10-year retention, low power consumption, fast reading and writing, complete compatibility with the CMOS process, and scaling ability [1-5]. However, the endurance characteristics are generally 10$^4$~10$^5$ [6-13], which is far from the application requirements of memory and computing in-memory application (>10$^{14}$) [14]. Therefore, the study of endurance characteristics has attracted widespread attention.

Manuscript received July 6, 2021. This work was supported by the National Natural Science Foundation of China under Grant No. 61904199 and 61904193, and in part by the Open Research Project Fund of State Key Laboratory of ASIC and System under Grant No. KVH1233021. (Corresponding author: Hao Xu, Jinjuan Xiang)

Fengbin Tian, Shujing Zhao, Hao Xu, Jinjuan Xiang, Tingting Li, Wenjuan Xiong, Jiahui Duan, Junshuai Chai, Kai Han, Xiaolei Wang, Wenwu Wang, and Tianchun Ye are with Key Laboratory of Microelectronics Devices and Integrated Technology, Institute of microelectronics, Chinese academy of sciences, Beijing 100029, China. The authors are also with University of Chinese Academy of Sciences, Beijing 100049, China (xuhao@ime.ac.cn; xiangjinjuan@ime.ac.cn).

Kai Han is with Weifang University, Shandong 261061, China.

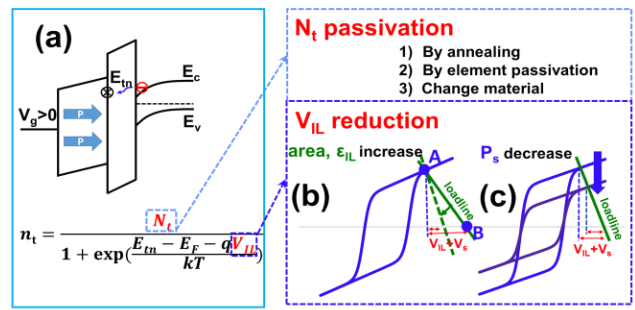

Fig.1. (a) Energy band diagram of MFIS gate stack showing the charge trapping behavior. The n$_t$ means trapped charge density. (b) and (c) Ferroelectric hysteresis loop and loadline showing the principle of decreasing voltage drop across the interlayer.

Significant charge trapping/de-trapping is widely accepted as the origin of endurance fatigue. Thus, effective suppression of charge trapping is rather important to improve endurance characteristics [15-19]. Fig. 1 shows the charge trapping mechanism of the MFIS (TiN/Hf$_x$Zr$_{1-x}$O$_2$/interlayer/Si) gate stack. Two parameters determine the charge trapping behavior, i.e., trap density ($N_t$) and tunneling barrier. Thus, the current method of improving endurance can be summarized as the following two kinds.

The first method is to reduce the $N_t$. The self-heating effect [7] and high-pressure hydrogen annealing [8] can reduce the $N_t$. In addition, the introduction of the N element in the SiO$_2$ interlayer can also effectively suppress trap generation [9, 20, 21]. However, these works claim the trap passivation by measuring the endurance characteristics, but not *directly* measuring the charge trapping behavior.

The second method is to reduce voltage drop on the interlayer ($V_{IL}$), including i) increasing dielectric constant of interlayer [9, 10, 22, 23], ii) decreasing the area ratio of the ferroelectric and interlayer capacitors [24], iii) eliminating interlayer [5, 25], iv) using Ω-Gate or recessed channel [26, 27], and v) reducing the spontaneous polarization ($P_s$) of ferroelectric [28-30]. Fig. 1(b) shows the ferroelectric hysteresis loop and loadline of the MFIS gate stack to explain the methods (i) and (ii). The intersection between the loadline and hysteresis loop is denoted as 'A'. The intersection between the loadline and abscissa is denoted as 'B'. Then the sum of voltage drop on the interlayer ($V_{IL}$) and surface potential ($V_s$) is given as the horizontal distance between 'A' and 'B'. Thus, the loadline will become steeper for the methods (i) and (ii). Then the distance decreases, which means that the $V_{IL}$ decreases at the same voltage drop across ferroelectric. Fig. 1(c) shows the ferroelectric hysteresis loop and loadline of the MFIS gate stack for the method (v). By reducing the $P_s$, the $V_{IL}$ and consequent interlayer electric field ($E_{IL}$) decrease at the same applied gate voltage. For the method of (v), theoretical predictions have been given [29], while there is no experimental study yet.

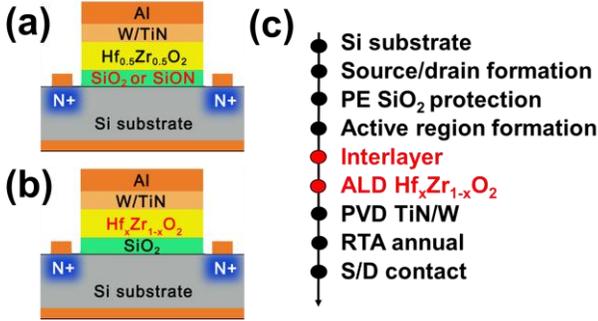

Fig. 2. The schematic of FeFET gate structure of (a) different interlayers and (b) different $Hf_xZr_{1-x}O_2$ materials, and (c) the fabrication process flow.

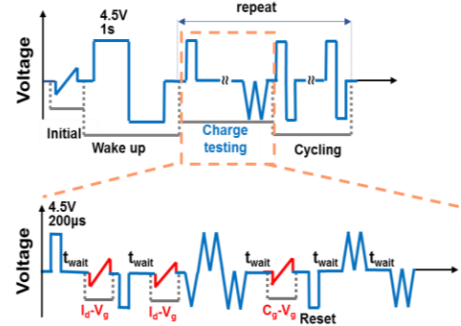

Fig. 3. The details of electrical measurement.

In this work, we experimentally study the impact of different interlayers and ferroelectric materials on charge trapping during the endurance fatigue of Si FeFET with $Hf_xZr_{1-x}O_2$/interlayer gate stack. For the different interlayer materials, we employ $SiO_2$ or SiON. For the impact of $P_s$ on endurance fatigue, we employ $Hf_xZr_{1-x}O_2$ layers by changing the Hf:Zr ratio. We also *directly* measure the charge trapping behavior.

## II. EXPERIMENTAL

### A. Device fabrication and measurements

The devices were fabricated by using a gate-last process. The silicon substrate used P-type silicon with a resistivity of about 10-12 Ω·cm. The source/drain was implanted using 40 keV energy with $4\times10^{15}$ cm$^{-2}$ dose of As ions. The gate structure was W/TiN/$Hf_xZr_{1-x}O_2$/interlayer/Si. The fabricated FeFET devices are summarized as shown in Fig. 2. Fig. 2(a) schematically shows FeFET with different interlayer materials. There are two kinds of interlayers. One is the $SiO_2$ interlayer, grown by ozone oxidation, and its thickness is about 0.7 nm. The other interlayer is SiON, grown by nitrogen plasma treatment of $SiO_2$, and its thickness is about 1.0 nm. Then same $Hf_{0.5}Zr_{0.5}O_2$ ferroelectric was used for different interlayers. The 9 nm $Hf_{0.5}Zr_{0.5}O_2$ was grown by ALD at 300 °C using tetrakis-(ethylmethylamino)-hafnium (TEMA-Hf) as Hf precursor, tetrakis-(ethylmethylamino)-zirconium (TEMA-Zr) as Zr precursor, and $H_2O$ as O source. Fig. 2(b) schematically shows FeFET with different ferroelectric materials. There are three kinds of $Hf_xZr_{1-x}O_2$ with x=0.75, 0.6, and 0.5. The interlayers are all 0.7 nm $SiO_2$. The $Hf_xZr_{1-x}O_2$ is 9 nm thick. The different Hf:Zr ratios were realized by changing the Hf and Zr precursor cycles. For all samples, the 10 nm TiN and 75 nm W were deposited by sputtering as the metal gate. Then the ferroelectric phase crystallization was achieved by 550 °C for 60 s in $N_2$ to form the orthorhombic phase. Then, the source/drain contacts were defined by lithography, and TiN/Al was used as contact metal. Finally, forming gas annealing at 450 °C in 5%-$H_2$/95%-$N_2$ was performed. In addition, the capacitor of TiN/$Hf_xZr_{1-x}O_2$/TiN was also fabricated with the same condition as above.

Fig. 3 shows the details of the electrical measurement we used in this work. We used Agilent B1500A and Radiant Precision LC ferroelectric tester to measure the transfer curve $I_d$-$V_g$, gate capacitance vs. gate voltage ($C_g$-$V_g$), the hysteresis loop, and PUND (Positive Up Negative Down). For MFM capacitors, a triangle wave of 3 V amplitude at 1 kHz was used for hysteresis loop measurement, and a square wave of 3 V amplitude at 1 kHz was used for the endurance characteristics test. For FeFET, AC small-signal measurement at 100 kHz was used for the $C_g$-$V_g$ curves test. A square pulse of ±4.5 V amplitude and 200 μs width was used for the endurance test. The PUND was measured using a triangular wave of 4.5 V amplitude at 200 μs pulse width. The threshold voltage was extracted by the linear extrapolation method. The interfacial density ($D_{it}$) was measured by the conductance method.

### B. Direct extraction of trapped charges

The direct measurement of trapped charges has been given in our previous work [19]. The main theory of this measurement method is described as follows. Generally, electrons are trapped into the gate stack after program (PGM) pulse, and here the trapped charge amount is denoted as $Q_{t,PGM}$. Holes are trapped after erase (ERS) pulse, and the trapped charge amount is denoted as $Q_{t,ERS}$. We experimentally measure the difference between the $Q_{t,PGM}$ and $Q_{t,ERS}$, i.e., $\Delta Q_t = Q_{t,PGM} - Q_{t,ERS}$ in this work, which represents the charge trapping behavior.

According to the charge neutrality condition, we can obtain

$$\Delta Q_t = -(\Delta Q_m + \Delta Q_{Si})$$

where the $\Delta Q_m$ is the charge change on the metal gate, and $\Delta Q_{Si}$ is the charge change on the substrate. For the $\Delta Q_m$, we can obtain it by integrating the external circuit gate current. For the substrate charges $\Delta Q_t$, it is equal to the difference of the $C_g$-$V_g$ curve integral from 0 to the respective threshold voltage ($V_{th}$) after positive and negative pulses, that is

$$\Delta Q_{Si} = \int_0^{V_{th,PGM}} C_{g,PGM}(V_g)dV_g - \int_0^{V_{th,ERS}} C_{g,ERS}(V_g)dV_g$$
$$= \int_{\Delta V_{th}}^{V_{th,ERS}} C_{g,ERS}(V_g)dV_g - \int_0^{V_{th,ERS}} C_{g,PGM}(V_g)dV_g$$
$$= -\int_0^{\Delta V_{th}} C_{g,ERS}(V_g)dV_g$$

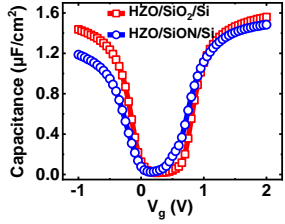

Fig. 4. The $C_g$-$V_g$ curves of FeFET with $SiO_2$ and SiON.

Thus, the trapped charges $\Delta Q_t$ can be obtained when the $\Delta Q_m$ and $\Delta Q_{Si}$ are obtained. The details of this method can be found in ref. [19].

## III. RESULTS AND DISCUSSION

### A. Different interlayers

Fig. 4 shows the $C_g$-$V_g$ curves of FeFET with $SiO_2$ and SiON. During measurement, the source, drain, and bulk were connected as a terminal, while the gate was the other terminal. The equivalent oxide thickness (EOT) of SiON and $SiO_2$ samples are 19.5 Å and 18.2 Å, respectively. Then the dielectric constant of SiON can be obtained as follows [23]

$$\kappa_{SiON} = \kappa_{SiO_2} \times \frac{t_{SiON}}{t_{SiO_2} - \delta_{EOT}}$$

where $\kappa_{SiON}$ and $\kappa_{SiO_2}$ are the dielectric constants of SiON and $SiO_2$, respectively; $t_{SiON}$ and $t_{SiO_2}$ represent the physical thicknesses of SiON and $SiO_2$, respectively; $\delta_{EOT}$ is the EOT difference between these two interlayers. Finally, the $\kappa_{SiON}$ is determined to be 4.7. This value is consistent with the reported results [31, 32].

Fig. 5(a) and (b) show the $I_d$-$V_g$ curves of $SiO_2$ and SiON samples after wake up, respectively. The memory window (MW) is nearly the same, i.e., 1.54 V and 1.56 V. Figs. 5(c)-(f) show the $I_d$-$V_g$ curves of $SiO_2$ and SiON samples during the endurance process. The subthreshold swing (SS) degrades during the endurance fatigue, indicating that the $D_{it}$ is increasing.

The $I_d$-$V_g$ degradation is different between the $SiO_2$ and SiON samples. The initial $I_{on}@V_g=V_{th}+1.0$ V of the $SiO_2$ sample is 3.11 μA/μm, and it decreases to 31.3% and 11.7% after the first PGM and ERS operation, respectively. In contrast, the initial $I_{on}@V_g=V_{th}+1.0$ V of the SiON sample is 3.14 μA/μm, and it decreases to 96.9% and 43.9% after the first PGM and ERS operation, respectively. Thus, the $I_{on}$ degradation is more severe for the $SiO_2$ sample. The physical origin of $I_{on}$ degradation after PGM/ERS operation is Coulomb scattering from increased trapped charges at $Hf_{0.5}Zr_{0.5}O_2$/interlayer interface and interfacial charges at interlayer/Si interface, which will be discussed later.

The $V_{th}$ can be extracted and the endurance characteristics of FeFET with different interlayers are shown in Fig. 6(a). The MW decreases with increasing the PGM/ERS cycle. The endurance of the SiON sample reaches $10^5$ cycles, which is 10 times larger than that of the $SiO_2$ sample. In addition, the initial MW of the SiON sample is 1.56 V. The above result is similar to the reported result [12].

We quantitatively investigate the charge trapping behavior. Fig. 6(b) shows the measured trapped charges into the gate stack as a function of the cycle number. It can be seen that the trapped charges of the SiON sample are always lower than that of the $SiO_2$ sample. In addition, the trapped charges of the SiON sample increase slower than that of the $SiO_2$ sample as the PGM/ERS cycle increases. Moreover, when the PGM/ERS cycle increases from $10^0$ to $10^4$, the trapped charges of SiON sample increase to 43.5 μC/cm², which is smaller than $SiO_2$ sample (136.5 μC/cm²). This indicates that more traps have been generated for the $SiO_2$ sample. Considering that the charge trapping/de-trapping occurs from the Si substrate into $Hf_xZr_{1-x}O_2$/interlayer interface, these results indicate that the introduction of N element into the $SiO_2$ interlayer can effectively suppress charge trapping and trap generation at $Hf_xZr_{1-x}O_2$/$SiO_2$ interface. It should be noted that it is the first time that the charge trapping behavior is experimentally extracted for the SiON sample and that the charge trapping behavior is directly verified to be effectively suppressed compared with the $SiO_2$ sample.

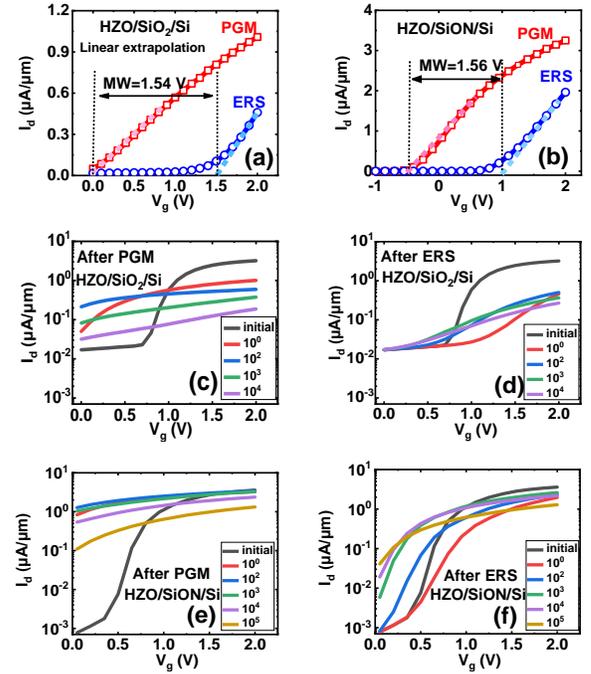

Fig. 5. (a) and (b) show the $I_d$-$V_g$ curves of $SiO_2$ and SiON samples after wake up, respectively. The $I_d$-$V_g$ curves during the endurance fatigue are given in (c) and (d) for the $SiO_2$ sample, and (e) and (f) for the SiON sample. Gate length/width is 5/150 μm for all the devices, and $V_d$=100 mV for $I_d$-$V_g$ measurement.

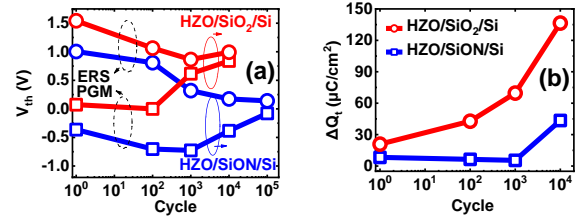

Fig. 6. (a) The endurance characteristics of FeFET with different interlayers and (b) the trapped charges during the cycling.

We next investigate the $D_{it}$. Fig. 7 shows the averaged $D_{it}$ across the bandgap after the erase operation as a function of the cycle number. It can be seen that the $D_{it}$ of the $SiO_2$ sample increases rapidly, from $10^{13}$ eV⁻¹·cm⁻² to $10^{14}$ eV⁻¹·cm⁻² as the PGM/ERS cycle increases from $10^0$ to $10^4$. Whereas the $D_{it}$ of the SiON sample is much lower, and it is nearly unchanged until the $10^3$ cycles. This proves that the introduction of the N element can effectively suppress the $D_{it}$ generation in FeFET. Nitrogen incorporation is well known to strengthen the $SiO_2$/Si interface by passivating dangling bonds and forming a stronger bond with Si [10]. Therefore, the introduction of the N element can effectively suppress trap generation both at the $Hf_xZr_{1-x}O_2/SiO_2$ and $SiO_2$/Si interfaces.

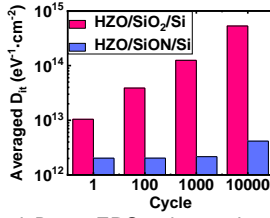

Fig. 7. The averaged $D_{it}$ vs. ERS pulse cycle of FeFET devices with different interlayers.

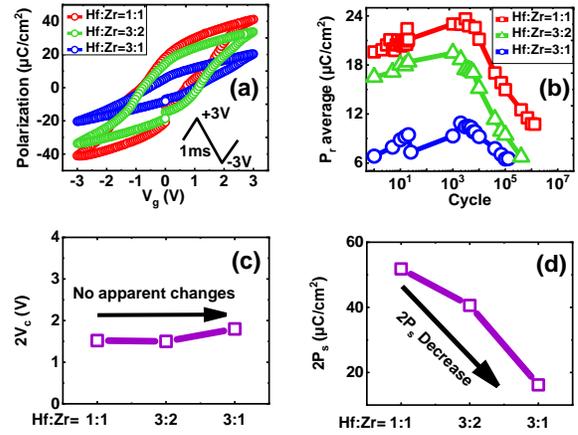

Fig. 8. (a) The initial P-V curves and (b) the fatigue characteristics of MFM capacitors with different Hf:Zr ratios. The $P_r$ means remanent polarization. (c) The initial $2V_c$ and (d) the initial $2P_s$ of MFM capacitors vs. different Hf:Zr ratios.

We discuss the energy position of traps at the $Hf_{0.5}Zr_{0.5}O_2/SiO_2$ interface. The endurance fatigue of $SiO_2$ and SiON samples is due to the $V_{th}$ increase after PGM and decrease after ERS. Considering that the trapped charges increase during endurance fatigue, we can conclude that the generated traps are localized both near the conduction band and valence band.

### B. $Hf_xZr_{1-x}O_2$ with different Hf:Zr ratios

We investigate the impact of $Hf_xZr_{1-x}O_2$ with different Hf:Zr ratios on the endurance characteristics. Firstly, we investigate the MFM capacitors. Fig. 8(a) shows initial polarization-voltage (*P-V*) curves of MFM capacitors. It can be seen that the Hf:Zr ratio effectively tunes the *P-V* curve. Fig. 8(b) shows endurance characteristics of the MFM capacitors with different Hf:Zr ratios. All the samples reach an endurance of $10^5 \sim 10^7$ cycles. The coercive voltage ($V_c$) and saturated spontaneous polarization ($P_s$) are summarized as shown in Fig. 8(c) and (d), respectively. It can be seen that for samples $Hf_{0.5}Zr_{0.5}O_2$, $Hf_{0.6}Zr_{0.4}O_2$, and $Hf_{0.75}Zr_{0.25}O_2$, the initial values of $2V_c$ are similar and about 1.5 V, while the $2P_s$ decreases from 51.8 $\mu C/cm^2$ to 16.2 $\mu C/cm^2$.

Secondly, we investigate the FeFET with different Hf:Zr ratios. Fig. 9 shows the initial $I_d$-$V_g$ curves, together with those after the $10^0$ PGM/ERS cycle. The $I_d$-$V_g$ degradation is observed. The initial $I_{on}@V_g=V_{th}+1.0$ V for all samples is ~3.0 $\mu A/\mu m$. After the first PGM and ERS operation, $I_{on}$ decreases to 31.4% and 14.3% for $Hf_{0.5}Zr_{0.5}O_2$ sample, 46.3% and 28.8% for $Hf_{0.6}Zr_{0.4}O_2$ sample, 81.4% and 79.5% for $Hf_{0.75}Zr_{0.25}O_2$ sample. The physical origin of $I_{on}$ degradation after PGM/ERS operation is Coulomb scattering from increased trapped charges and $D_{it}$. The subthreshold swing also degrades during the endurance fatigue, indicating that the $D_{it}$ is increasing.

The $V_{th}$ can be extracted and the MW is shown in Fig. 10. The endurance of all samples is ~$10^4$ cycles. After $10^4$ cycles, the gate stack breakdown occurs. Therefore, we use the MW degradation percent from $10^0$ to $10^3$ cycles to compare the endurance characteristics, as shown in Fig. 10(d). As the $P_s$ decrease from 25.9 $\mu C/cm^2$ of $Hf_{0.5}Zr_{0.5}O_2$ to 20.3 $\mu C/cm^2$ of $Hf_{0.6}Zr_{0.4}O_2$, the endurance characteristics improve. In contrast, when the $P_s$ further decreases from 20.3 $\mu C/cm^2$ of $Hf_{0.6}Zr_{0.4}O_2$ to 8.1 $\mu C/cm^2$ of $Hf_{0.75}Zr_{0.25}O_2$, the endurance characteristics degrade.

Then we quantitatively investigate the charge trapping behavior. Fig. 11(a) shows the measured trapped charges into the gate stack as a function of the cycle number. It can be seen that the trapped charges decrease in the sequence of $Hf_{0.5}Zr_{0.5}O_2$, $Hf_{0.6}Zr_{0.4}O_2$, and $Hf_{0.75}Zr_{0.25}O_2$, which is the same as the changing trend of $P_s$. Especially the charge trapping behavior nearly disappears for the $Hf_{0.75}Zr_{0.25}O_2$ sample. Therefore, decreasing the ferroelectric $P_s$ can effectively suppress the charge trapping.

However, as shown in Fig. 11(b), as the charge trapping effect becomes weaken, the endurance characteristics initially improve and then degrade. Moreover, it is worthy to note that the initial MW of the $Hf_{0.75}Zr_{0.25}O_2$ sample is nearly zero, and MW is all negative after $10^0$ PGM/ERS cycle, even though the $2E_c$ of $Hf_{0.75}Zr_{0.25}O_2$ sample is similar to others. We discuss the physical origin by analyzing the relationship among the polarization, charge trapping, and the MW as follows.

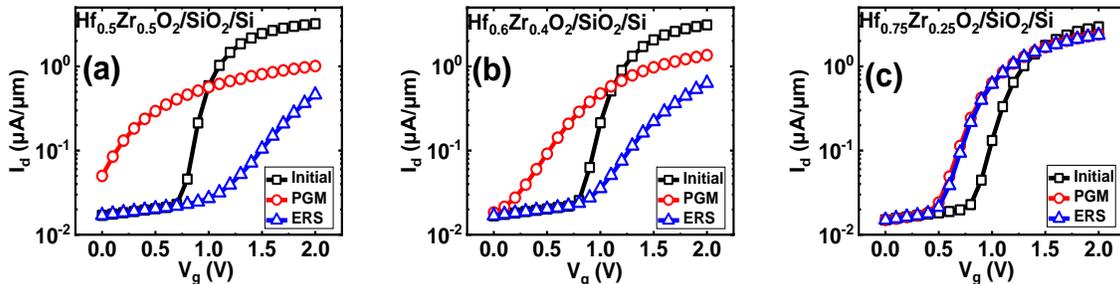

Fig. 9. The initial $I_d$-$V_g$ curves and the $I_d$-$V_g$ curves after the 1 cycle of PGM/ERS operation for FeFET with different Hf:Zr. Gate length/width is 5/150 μm for all the devices, and $V_d$=100 mV for $I_d$-$V_g$ measurement.

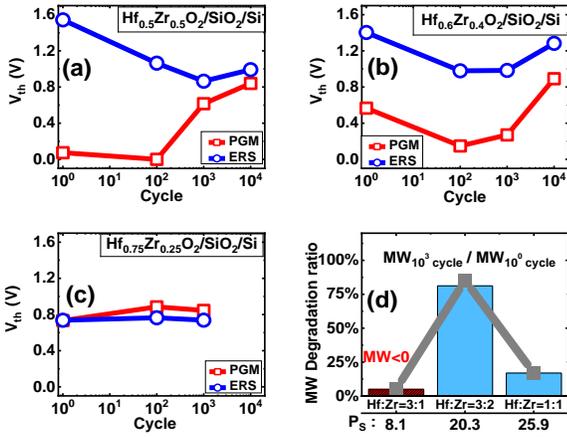

Fig. 10. (a)-(c) The endurance characteristics of FeFET with different Hf:Zr ratios. (d) MW degradation ratio ($10^3$ cycles to $10^0$ cycles) vs. Hf:Zr ratio.

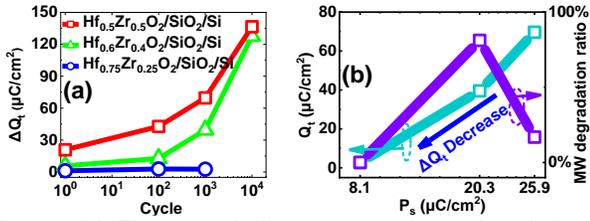

Fig. 11. (a) The trapped charges vs. cycle number during the endurance process for FeFET with different Hf:Zr ratios. (b) The $\Delta Q_t$ and endurance characteristics in $10^3$ cycles vs. the $P_s$.

Firstly, we discuss the endurance improvement from $Hf_{0.5}Zr_{0.5}O_2$ to $Hf_{0.6}Zr_{0.4}O_2$ samples. Fig. 12 shows the ferroelectric hysteresis loop and loadline of the MFIS gate stack for the three samples. All the curve numbers are labled as 1, 2, and 3 for the $Hf_{0.5}Zr_{0.5}O_2$ sample ($P_s$=25.9 μC/cm$^2$), the $Hf_{0.6}Zr_{0.4}O_2$ sample ($P_s$=20.3 μC/cm$^2$), and the $Hf_{0.75}Zr_{0.25}O_2$ sample ($P_s$=8.1 μC/cm$^2$), respectively. By reducing the $P_s$, the $V_{IL}$ and consequent $E_{IL}$ decrease at the same applied gate voltage (from $V_{IL,1}$ to $V_{IL,2}$). Thus, the endurance improves. In addition, the MW decreases from $Hf_{0.5}Zr_{0.5}O_2$ to $Hf_{0.6}Zr_{0.4}O_2$ sample as shown in Fig. 10. This is explained as follows. At $V_g$=$V_{th}$, the substrate charge is defined as $Q_{Si,th}$. For the case without charge trapping, the charges in the metal gate corresponding to $V_{th}$ are equal to $Q_{Si,th}$, which is shown as the dash lined 'TH0' in Fig. 12(a). When charge trapping appears, at $V_g$=$V_{th}$, the charges in the metal gate corresponding to $V_{th}$ are equal to $Q_{Si,th}+Q_t$, which is shown as the dashed line 'TH1'.

Then the MW can be determined to be the distance between the intersections of the hysteresis loop and the line 'TH1'. When ferroelectric material changes from $Hf_{0.5}Zr_{0.5}O_2$ to $Hf_{0.6}Zr_{0.4}O_2$, the $P_s$ reduces and the hysteresis loop changes from the line 'HL1' to the line 'HL2'. The trapped charges slightly reduce and the line representing $V_{th}$ changes from the line 'TH1' to the line 'TH2'. Because the $P_s$ reduction is larger than the trapped charge reduction, the line 'HL2' drops more than the line 'TH2' as shown in Fig. 12(b). Therefore, the MW decreases.

Secondly, we discuss the MW of ~0.02 V for $Hf_{0.75}Zr_{0.25}O_2$ sample ($P_s$=8.1 μC/cm$^2$). Due to the small $P_s$, the intersections between the hysteresis loop line 'HL3' and line 'TH3' can be localized in the upper right section as shown in Fig. 12(c). Thus, the MW can be significantly reduced.

From the above discussion, it can be concluded that the reduction of $P_s$ could improve endurance characteristics. On the contract, it can also reduce the MW. The above two points are determined by the competition between $P_s$ reduction and charge trapping induced by it. Therefore, the impact of $P_s$ reduction on endurance fatigue needs careful consideration.

IV. CONCLUSION

We study the impact of different interlayers and ferroelectric materials on charge trapping during the endurance fatigue of Si FeFET with $Hf_xZr_{1-x}O_2$/interlayer gate stack and analyze the endurance degradation mechanism caused by the charge trapping. We directly extract the charge trapping behavior for $SiO_2$ and SiON interlayer for the first time. The introduction of the N element in the $SiO_2$ interlayer effectively suppresses the charge trapping and defect generation and improves the endurance characteristics. By reducing the saturated spontaneous polarization of ferroelectric, the charge trapping can be effectively suppressed, and the endurance characteristics can also be improved. However, it can also cause MW degradation. Our work helps design the MFIS gate stack.

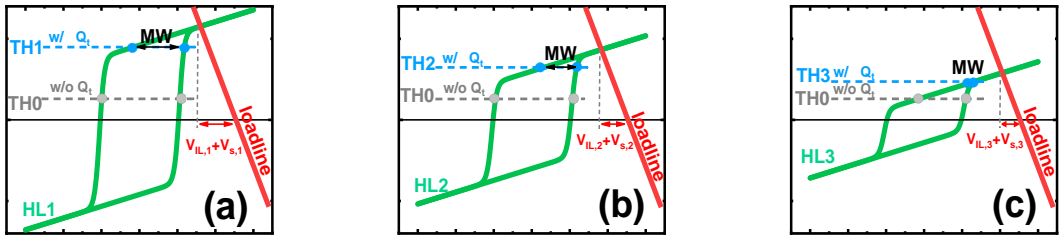

Fig.12. Ferroelectric hysteresis loop and Loadline simulation model of memory window and endurance characteristics changing with the ferroelectric reduction.